\documentclass[12pt]{article}

\usepackage{times}
\usepackage[margin=1in]{geometry}
\usepackage{setspace}
\usepackage{lineno}
\usepackage{graphicx}
\usepackage{amsmath, amssymb}
\usepackage{authblk}
\usepackage{caption}
\usepackage{abstract}
\usepackage[numbers,sort&compress]{natbib}
\usepackage{comment}

\usepackage{graphicx}
\usepackage{bm}
\usepackage{float}

\usepackage[dvipsnames]{xcolor}
\usepackage[colorlinks=true, allcolors=blue]{hyperref}
\newcommand{\OS}[1]{}

\title{Finite Vertical Windows: Seeing Only Part of the Picture in Rotating Turbulence}
\author{Omri Shaltiel, Eran Sharon}

\begin{document}
\maketitle

\begin{abstract}
We report high-resolution measurements of three-dimensional (3D) turbulence in a rapidly rotating fluid. By decomposing the velocity field into a vertically averaged component and a three-dimensional residual, we show that each dominates distinct frequency ranges: the quasi-2D component at low frequencies and the 3D component at higher ones. This separation is not intrinsic to the flow but strongly depends on the finite vertical span of the measurements. As the vertical scan range increases, the apparent crossover between 2D and 3D-dominated regimes shifts systematically, revealing that the commonly assumed partition is strongly shaped by measurement limits. These findings call into question the usage of the concept of pure 2D manifold, in the theoretical description of rotating turbulence and highlight the need for frameworks that account for resolution-dependent parts of the flow and the coupling between wave-like and vortex-like motions.

\end{abstract}

\section{Introduction}
Understanding rotating turbulence is a fundamental and enduring challenge in fluid dynamics, with profound implications for a variety of geophysical and astrophysical phenomena. The influence of rotation fundamentally alters turbulent flow, breaking the isotropy inherent to classical turbulence and introducing strong anisotropy into the energy transfer across scales. In such systems, two dynamically distinct regimes typically emerge. At large scales, coherent, long lived quasi-2D structures align with the rotation axis \cite{alexakis2018cascades, biferale2016coherent,godeferd2015structure}. At smaller scales, the dynamics are governed by three-dimensional inertial wave interactions \cite{davidson2013turbulence, galtier2003weak, cortet2020quantitative, cortet2020shortcut, cortetcampagne2015disentangling, shaltiel2025confirming}. Understanding the relations between these two "components" of the flow is a central task in the theoretical modeling of rotating turbulence.

The governing equations of rotating turbulence are the incompressible Navier-Stokes equations expressed in a rotating reference frame. Introducing a constant angular velocity vector $\bm{\Omega}$, these equations become \cite{greenspan1968theory}:
\begin{equation}
    \frac{\partial \boldsymbol{v}}{\partial t} + (\boldsymbol{v} \cdot \nabla) \boldsymbol{v} + 2 \boldsymbol{\Omega} \times \boldsymbol{v} = -\frac{1}{\rho} \nabla p + \nu \nabla^2 \boldsymbol{v},
\end{equation}

where $\bm{v}$ is the velocity field, $p$ is the pressure, $\rho$ is the density, $\nu$ is the kinematic viscosity, $\bm{\Omega}$ is the rotation vector around the vertical axis, which we denote as $\hat{z}$, so $\bm{\Omega}=\Omega \hat{z}$. Two key dimensionless parameters describe rotating turbulence: the Reynolds number, $\mathrm{Re}=UL/\nu$, which quantifies inertial relative to viscous effects, and the Rossby number, $\mathrm{Ro}=U/(2\Omega L)$, which measures inertial relative to Coriolis forces. \OS{where $U$ and $L$ are characteristic velocity and length scales.} Rotating turbulent flows typically exhibit large $Re$ and small $Ro$, implying strong rotation and a broad inertial range.

One hallmark of rotating turbulence at large scales is the formation of coherent vortices aligned along the axis of rotation. These structures result from an inverse energy cascade, an inherently antidiffusive process wherein energy accumulates at progressively larger scales rather than dissipating to smaller scales. This inverse cascade has been extensively documented, both experimentally\cite{godeferd2015structure,cortetcampagne2015disentangling, yarom2013experimental,shaltiel2024direct} and numerically \cite{mininni2010rotating, deusebio2014dimensional, biferale2016coherent}, and is commonly interpreted as evidence of the system's effective dimensional reduction from 3D to quasi-2D dynamics~\cite{smith1999transfer, yarom2013experimental, buzzicotti2018inverse, campagne2014direct,biferale2016coherent}.

In earlier work ~\cite{shaltiel2024direct}, we showed that despite this apparent \OS{two-dimensionalization, the large-scale structures in rotating turbulence are underlain by low frequency inertial waves}. We demonstrated through detailed experimental measurements that these seemingly 2D vortices are, in fact, composed of inertial wave modes characterized by near-zero frequency and low vertical wavenumber $k_z$. At smaller scales, the dynamics transition to a regime governed by weak inertial wave interactions described effectively by Weak Wave Turbulence (WWT) theory \cite{shaltiel2025confirming, cortet2020quantitative}. WWT predicts a strongly anisotropic forward cascade with an energy spectrum computed by Galtier ~\cite{galtier2003weak}:
\begin{equation}
    E(k_r, k_z) \propto \sqrt{\epsilon \Omega} k_r^{-5/2} k_z^{-1/2},
    \label{eq:GaltierSpectrum}
\end{equation}
where $\epsilon$ denotes the energy injection rate, \OS{$k_r=\sqrt{k_x^2+k_y^2}$ denotes the horizontal (radial) wavenumber, and $k_z$ denotes the vertical wavenumber}.

While the weakly nonlinear inertial wave field provides a natural framework for describing small‐scale dynamics, it is well established that the quasi-2D component dominates the total energy of rotating turbulence. In many theoretical and experimental works~\cite{smith1999transfer,buzzicotti2018inverse,alexakis2018cascades,sen2012anisotropy,baroud2003scaling,lamriben2011direct,campagne2014direct,di2016quantifying,le2020near,cambon1997energy,bellet2006wave,gallet2015exact,nazarenko2011critical} considerable attention has therefore been devoted to understanding how energy is transferred from the wave field into the 2D component. This coupling between three‐dimensional inertial waves and 2D modes remains one of the central open questions in the theory of rotating turbulence. Specifically, it raises the question of how nonlinear interactions enable energy initially injected into the wave field to feed and sustain the 2D structures.

In the present study, we directly address the subtleties introduced by the finite vertical measurement range and its implications for resolving the vanishing-$k_z$ limit. Employing high-resolution 3D2C Particle Image Velocimetry (PIV) measurements of steady-state rotating turbulence, we investigate how finite vertical resolution impacts the decomposition between the quasi-2D and wave parts of the flow.

We show that the decomposition into quasi-2D and 3D components is set by the vertical measurement range rather than by the flow dynamics alone. Modes with very small but nonzero $k_z$ are inevitably included in the quasi-2D component. As a result, when increasing the measurement scale more energy is included in the "3D" component. In our experimental system, extrapolating this trend indicates that the 3D part of the flow is as energetic as the 2D component.

\section{Experimental System}
\begin{figure}[h]
    \centering
    \includegraphics[width=0.5\textwidth]{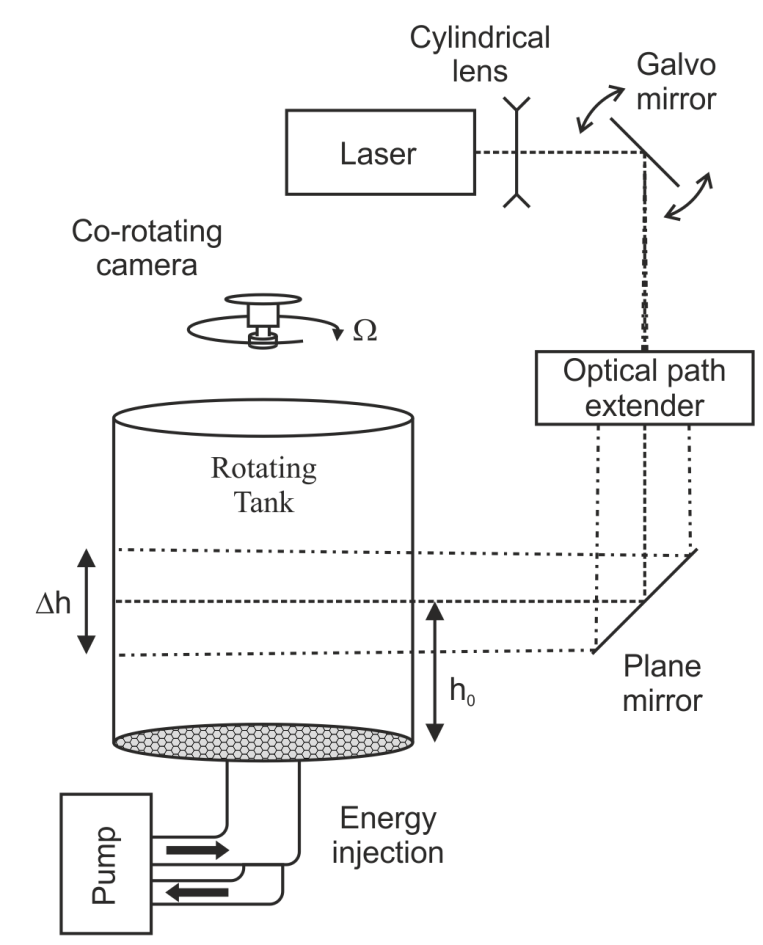}
    \caption{Schematic of the rotating-tank apparatus. A horizontal laser sheet, scanned vertically by a galvanometric mirror, illuminates seeding particles while a co-rotating high-speed camera records the flow.}
    \label{fig:experimental_system}
\end{figure}

Our experiment begins with a transparent acrylic cylinder (diameter $80~\text{cm}$, height $90~\text{cm}$) mounted on a computer-controlled turntable. A schematic of the apparatus is shown in Fig.~\ref{fig:experimental_system}. The tank rotates at a constant rate $\Omega$ up to $2~\text{Hz}$ around the vertical axis, $\hat{\boldsymbol{z}}$. The tank is filled with water and sealed with a rigid, transparent lid that flattens the free surface while offering a clear optical path to the interior.

We drive turbulence from the bottom. A hexagonal array of 248 silicone inlet tubes ($0.76~\text{mm}$ diameter, $11~\text{cm}$ length) alternates with 73 outlet ports ($6~\text{mm}$ diameter), forming a closed loop. A positive-displacement pump pushes the water through the loop providing steady forcing and a statistically steady flow.

The water is seeded with neutrally buoyant polyamide PIV particles (spheres $50~\mu\text{m}$). A $532~\text{nm}$ horizontal laser sheet illuminates the flow. The laser sheet is scanned vertically over a range $\Delta h \approx 24~\text{cm}$ using a galvanometric mirror that is swept over 30 steps. One full sweep takes $46.7~\text{ms}$.

A co-rotating high-speed camera records at $750~\text{fps}$ while the galvanometric mirror scans from $h_0-\Delta h/2$ upward until $h_0+\Delta h/2$ in synchrony with the shutter before resetting and repeating.

Images are processed using an in-house MATLAB implementation 3D2C PIV, based on OpenPIV-MATLAB \cite{ben2020openpiv}, adapted to our scanning protocol and accelerated on a GPU using MATLAB's gpu acceleration. This process yields a horizontal resolution of $0.29~\text{cm}$. Combined with the vertical step, the reconstructed volume is $77\times77\times30$ voxels at $21.4~\text{Hz}$. The algorithm results in the well resolved in-plane velocity components, $\boldsymbol{v} =\boldsymbol{v}_{\perp}(x,y,z,t)$. \OS{Throughout this paper, all reported energies and spectra are computed from the measured in-plane velocity $\boldsymbol{v}_\perp$.}

Each experiment began by spinning up the system, allowing it to reach steady rotation\OS{. The forcing was then switched on, for an additional $5~\text{min}$ to exclude the transient following forcing onset. Then, at statistical steady state we recorded $45~\text{s}$ of data}. Across all runs, the control parameters satisfy $\mathrm{Ro} \ll 1$ and the Reynolds number remained $\mathrm{Re} \gg 1$.

\section{Results}
\subsection{Spatio-temporal Decomposition of the Flow}

\begin{figure}[t]
    \includegraphics{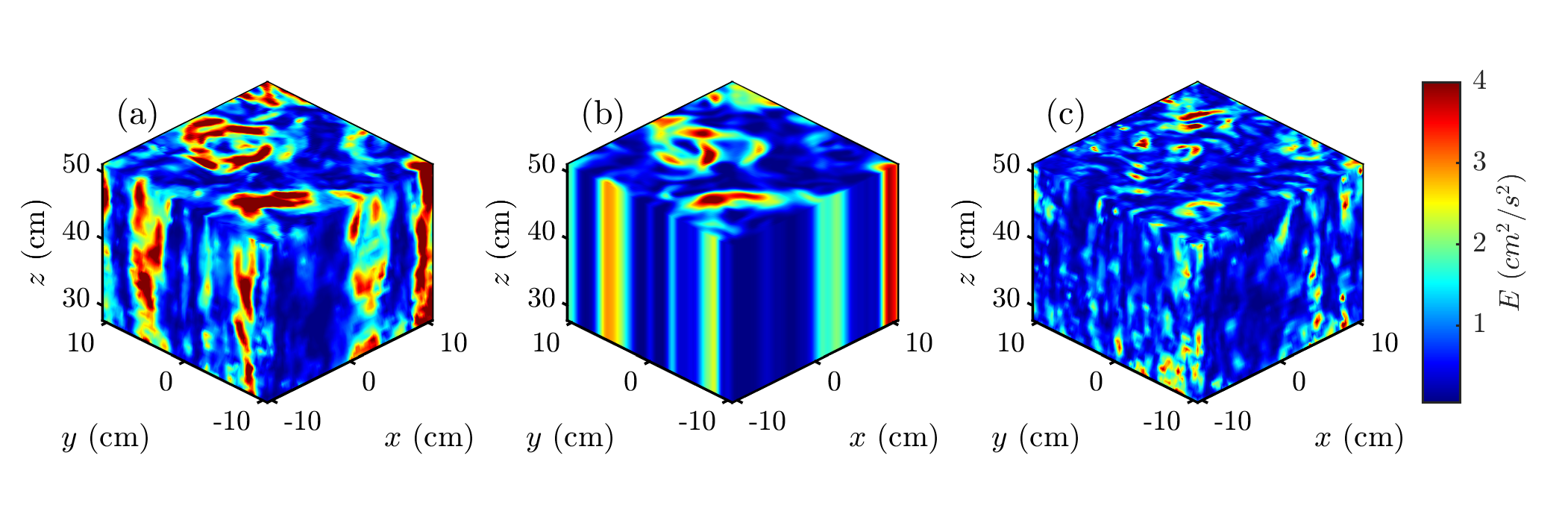}
    \includegraphics{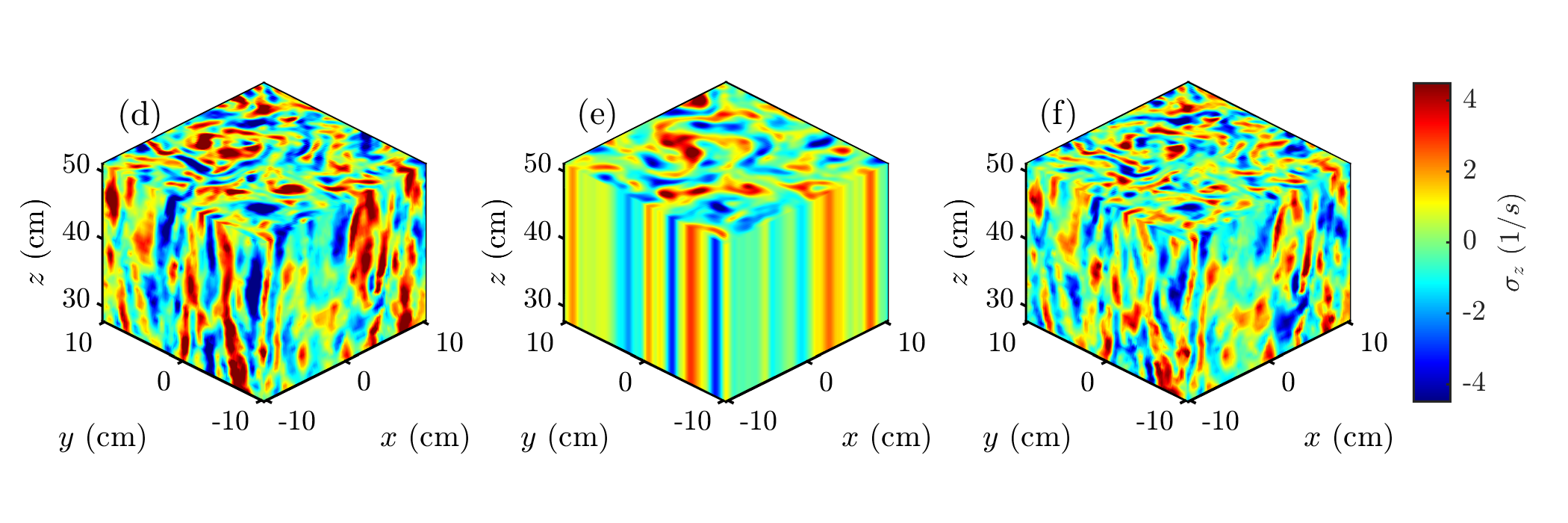}

   \caption{
    Visualization of the flow decomposition into quasi-2D and 3D components.
    (a–c) Instantaneous kinetic energy density $E$ and (d–f) vertical vorticity $\omega_z$ in a rotating turbulent flow, shown for: the full velocity field (left), its vertical average $\boldsymbol{v}_{2D}$ (center), and the residual field $\boldsymbol{v}_{3D}$ (right).  
    Data correspond to $\mathrm{Re}\approx 3000$ and $\mathrm{Ro}\approx 0.02$.  
    The full field exhibits broadband anisotropic turbulence with vertically coherent columns.  
    The quasi-2D component retains disordered, large-scale vortices that persist over many rotation periods, while the 3D component highlights smaller-scale, vertically modulated fluctuations with anisotropic vertical correlations.
    }

    \label{fig:flow-decomposition}
\end{figure}

We begin by examining the instantaneous kinetic energy density field shown in Figure~\ref{fig:flow-decomposition}(a), taken from an experiment with $\Omega = 4\pi~\mathrm{rad/s}$. The energy field consists of vertically elongated, columnar structures that persist throughout the measurement volume. These coherent columns are aligned with the rotation axis and exhibit vertical translation symmetry across the measurement volume, consistent with the formation of quasi-2D vortices. Their spatial extent along $\hat{z}$ spans the full height of the measured domain, suggesting a strong vertical correlation indicative of 2D dynamics.

In horizontal cross-sections, energy is distributed across a range of scales. Large, slowly evolving vortices dominate the horizontal organization of the quasi-2D dynamics, while smaller, rapidly varying fluctuations correspond to three-dimensional motions. The coexistence of large, coherent structures and small-scale fluctuations reflects the spectral heterogeneity of the flow, motivating a decomposition of the flow into dynamically distinct components.

\subsection{Flow Decomposition into Quasi-2D and 3D Components}

To isolate the quasi-2D component of the flow from three-dimensional components, we decompose the velocity field into a vertically averaged part $\boldsymbol{v}_{2D}(x,y,t)$ and a residual field $\boldsymbol{v}_{3D}(x,y,z,t)$. The vertically averaged component is defined as
\begin{equation}
    \boldsymbol{v}_{2D}(x,y,t) = \frac{1}{\Delta h} \int_{h_0-\Delta h/2}^{h_0+\Delta h/2} \boldsymbol{v}(x,y,z,t)\mathrm{d}z,
    \label{eq:vertical_averaging}
\end{equation}
with the residual
\begin{equation}
    \boldsymbol{v}_{3D}(x,y,z,t) = \boldsymbol{v}(x,y,z,t) - \boldsymbol{v}_{2D}(x,y,t).
\end{equation}
For a finite measurement height $\Delta h$, the vertical averaging in Eq.~\ref{eq:vertical_averaging} does not isolate a strictly $k_z = 0$ mode, but acts as a low-pass filter in $k_z$, retaining modes with $|k_z| \lesssim 2\pi/\Delta h$.

This decomposition is commonly interpreted as separating the $k_z = 0$ mode from the $k_z \neq 0$ modes of the flow \cite{shaltiel2025confirming, buzzicotti2018inverse, cortetcampagne2015disentangling, alexakis2023quasi}. Under the assumption of an infinitely extended horizontal system, Scott \cite{scott2014wave} showed analytically that such a decomposition cleanly separates a two-dimensional vortical component from a three-dimensional inertial-wave field when both evolve separately. 
Here we show experimentally that rotating turbulence exhibits behavior consistent with this separation, while also revealing its dependence on the finite measurement volume. 

A representative snapshot of this decomposition is shown in Fig.~\ref{fig:flow-decomposition}. The full flow field (left panels) displays anisotropic turbulence characterized by vertically coherent columns and broadband horizontal fluctuations. The vertically averaged field $\boldsymbol{v}_{2D}$ (center panels) retains the slowly evolving, horizontally organized vortices that dominate the kinetic energy. \OS{These structures evolve slowly compared to the rotation period, while still drifting, deforming, and interacting, and they exhibit characteristic features consistent with quasi-2D dynamics.} In contrast, the residual $\boldsymbol{v}_{3D}$ field (right panels) is composed of smaller-scale fluctuations with pronounced vertical modulation, consistent with inertial waves dynamics.

\subsection{Temporal Energy Spectra of the Total and Decomposed Flow}

The temporal energy spectrum, $E(\omega)$, captures how energy is distributed in frequency\OS{ (temporal angular frequency $\omega$)} and therefore directly reflects the orientation statistics of inertial waves through their dispersion relation $\omega = \pm 2\Omega \cos\theta$\OS{, with $\theta$ the angle between $\boldsymbol{k}$ and the rotation axis $\boldsymbol{\Omega}$}.
To quantify the temporal organization of energy, we compute the one-dimensional temporal energy spectrum:
\begin{equation}
    E(\omega) = \frac{1}{2TV} \int |\tilde{\boldsymbol{v}}(\omega)|^{2}~\mathrm{d}V,
\end{equation}
where $\tilde{\boldsymbol{v}}(\omega)$ is the temporal Fourier transform of the velocity field, $V$ is the measured volume, and $T$ is the total measurement duration. 

\begin{figure}[t]
    \centering
    \includegraphics[width=8cm]{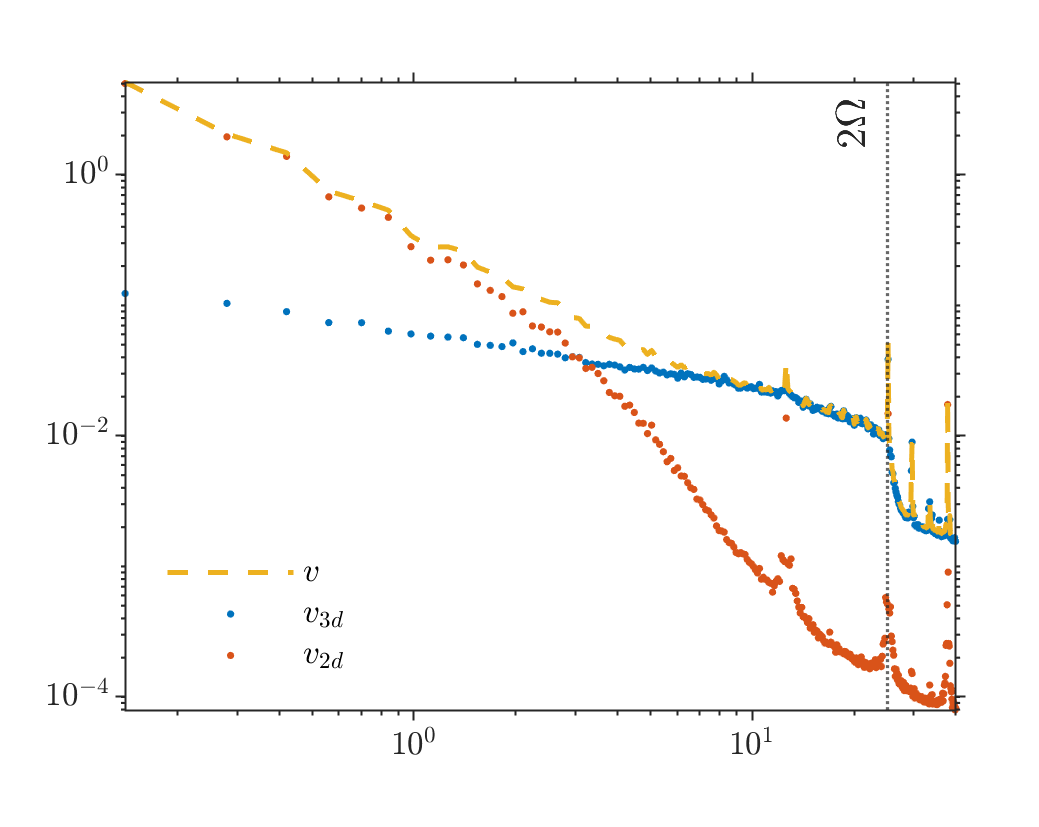}
    \includegraphics[width=8cm]{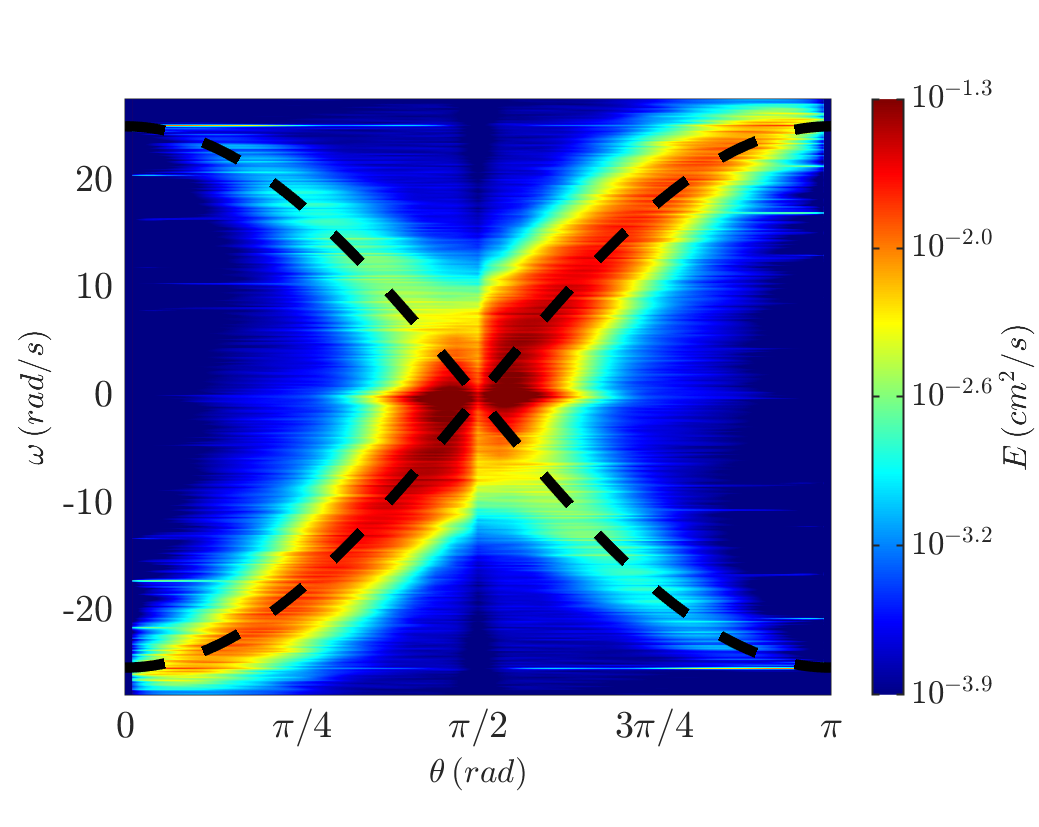}
    \caption{
    Left: Temporal energy spectra of the full velocity field (dashed blue), quasi-2D component $\bm{v}_{2D}$ (yellow), and 3D component $\bm{v}_{3D}$ (red). The quasi-2D contribution dominates at low frequencies, while the 3D component dominates at high frequencies and exhibits a sharp cutoff near $\omega = 2\Omega$, consistent with inertial-wave dispersion relation.  \OS{Right: The $(\theta,\omega)$ panel shows the energy distribution of the residual field $\boldsymbol{v}_{3D}$. The dashed curve marks the inertial-wave dispersion relation. the data is the same dataset as in Fig. 2.} }
    \label{fig:energy_time_spectrum_decomp}
\end{figure}

The temporal energy spectra of the total velocity field and its quasi-2D and 3D components are shown in Fig.~\ref{fig:energy_time_spectrum_decomp}. At low frequencies, the spectrum is dominated by the quasi-2D component, which follows an energy scaling close to $E_{2D}(\omega)\sim\omega^{-5/3}$. This regime reflects the slow evolution of large-scale vortical structures. A similar scaling was observed experimentally in rotating turbulence when coherent vortices were directly generated in a rotating tank~\cite{cortetcampagne2015disentangling}.

At higher frequencies, the three-dimensional component $\boldsymbol{v}_{3D}$ becomes dominant. Its spectrum exhibits a sharp cutoff near $\omega = 2\Omega$, consistent with the upper frequency bound imposed by the inertial-wave dispersion relation.
\OS{To further test the wave interpretation of the residual field, Fig.~~\ref{fig:energy_time_spectrum_decomp} (Right Panel) also shows the energy distribution of $\boldsymbol{v}_{3D}$ in the $(\theta,\omega)$ plane. The energy is concentrated in the inertial-wave band, remains bounded by $\omega \le 2\Omega$, and follows the inertial-wave dispersion relation. This is consistent with interpreting the separated $\boldsymbol{v}_{3D}$ field as wave dominated. The construction of the $(\theta,\omega)$ representation is described in the SI \cite{supp_material}.}

\begin{figure}[t]
    \centering
    \includegraphics[width=0.48\textwidth]{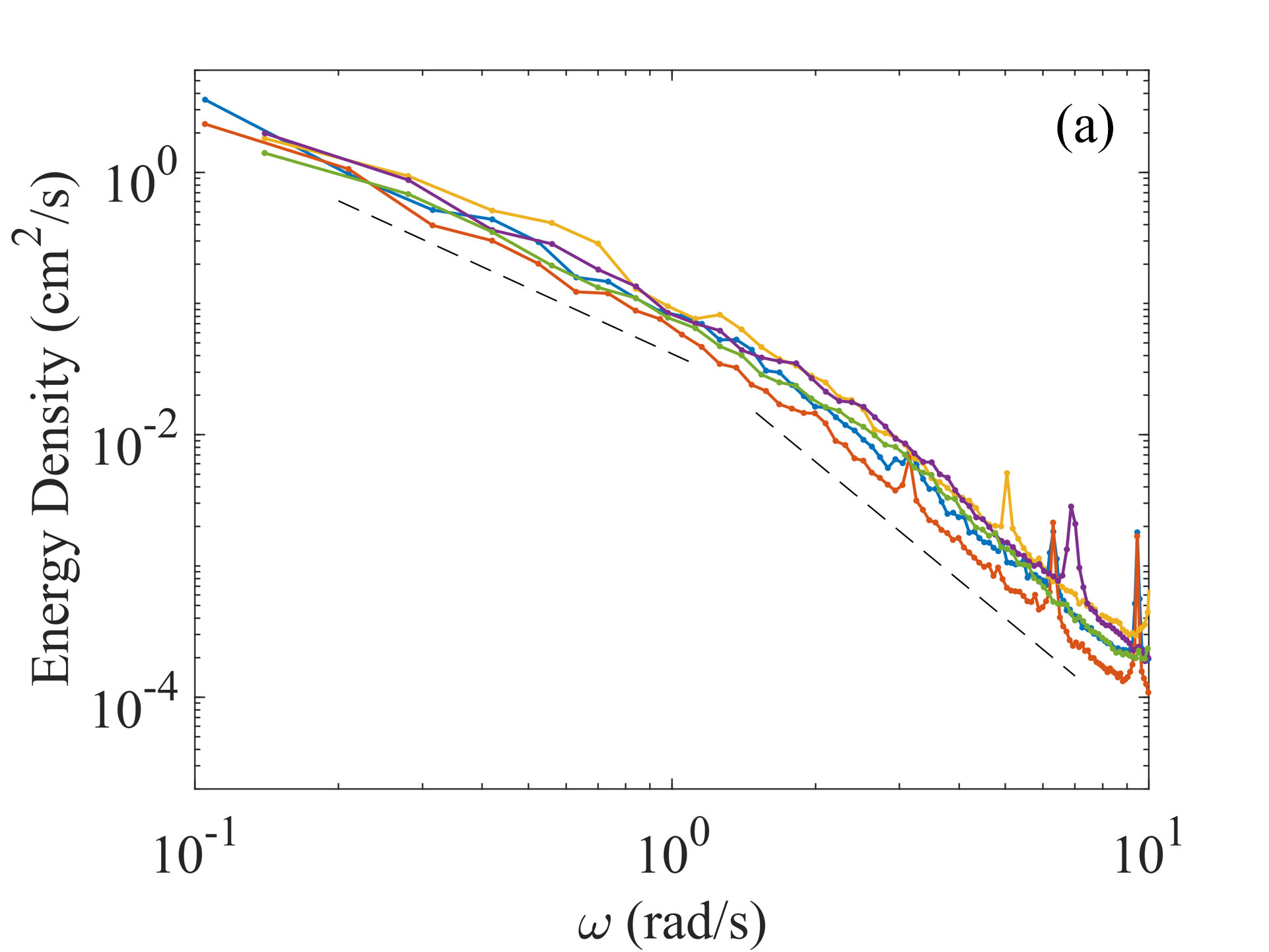}
    \includegraphics[width=0.48\textwidth]{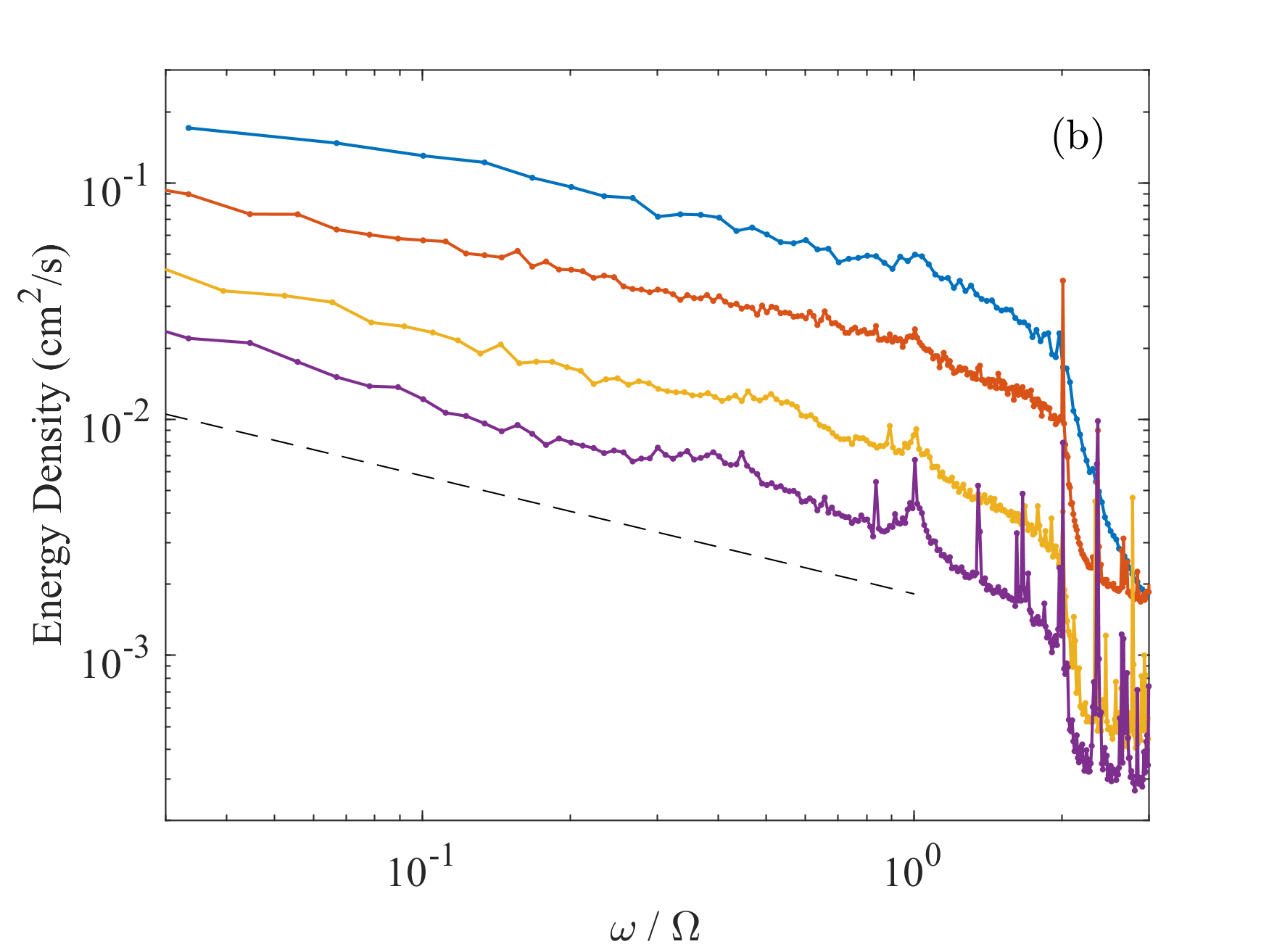}
    \caption{ Temporal energy spectra of (a) the quasi-2D field, $E_{2D}(\omega)$, and (b) the residual field, $E_{3D}(\omega)$.
    Colors indicate the rotation rate and forcing level (Table~S1): blue, $\Omega=0.5\times 2\pi~\mathrm{rad/s}$ (baseline forcing);
    green, $\Omega=1.0\times 2\pi~\mathrm{rad/s}$ (stronger forcing, $1.5\times$ baseline);
    yellow, $\Omega=1.7\times 2\pi~\mathrm{rad/s}$ (baseline forcing);
    red, $\Omega=2.0\times 2\pi~\mathrm{rad/s}$ (stronger forcing, $1.5\times$ baseline);
    purple, $\Omega=2.0\times 2\pi~\mathrm{rad/s}$ (baseline forcing).
    Dashed lines indicate the reference scalings: in (a) $E_{2D}(\omega)\sim \omega^{-5/3}$ at low frequencies and $E_{2D}(\omega)\sim \omega^{-3}$ at higher frequencies; in (b) $E_{3D}(\omega)\sim \omega^{-1/2}$.    
    }
    \label{fig:decomp_spectra}
\end{figure}

To compare the temporal behavior of the slow and fast components, we examine the spectra of the quasi-2D and residual (3D) velocity fields for different rotation rates (Figure~\ref{fig:decomp_spectra}).  
Panel~(a) shows the quasi-2D spectra $E_{2D}(\omega)$, which display two power-law regimes: $E_{2D}\propto\omega^{-5/3}$ at low frequencies and $E_{2D}\propto\omega^{-3}$ at higher frequencies.  
These scalings remain nearly unchanged across the explored range of $\Omega$.
These scaling regimes resemble the inverse-energy-cascade and forward-enstrophy-cascade scalings in 2D turbulence. Such a correspondence between temporal and spatial scalings could arise if a Taylor-type frozen-field hypothesis holds, linking frequency and wavenumber through $\omega \sim U k$~\cite{taylor1938spectrum,pecseli2022Taylorapplicability}. However, as far as we can tell, the standard conditions for this hypothesis, in particular the presence of a strong mean flow, are not fulfilled in our experiments. We also note that an approximate $k^{-5/3}$ scaling in horizontal wavenumber spectra was reported extensively in previous rotating-turbulence experiments and measurements, including Refs.~\cite{yarom2013experimental,yarom2014experimental,shaltiel2024direct}. By contrast, the forward-range spatial spectrum is typically steeper, but in our data it does not appear as a clear power-law range.


Panel~(b) presents the temporal energy spectrum of ${v}_{3D}$, the residual field  $E_{3D}(\omega)$.  
The high-frequency range follows $E_{3D}(\omega)\propto\omega^{-1/2}$, consistent with the prediction of Weak Wave Turbulence theory for inertial waves (See SI~\cite{supp_material} and~\cite{galtier2003weak, shaltiel2025confirming}). This observation highlights the distinct temporal signature of the wave-dominated component compared with the slowly evolving quasi-2D flow.

\subsection{Dependence of Spectral Separation on Vertical Resolution}

After showing the distinct nature of the 2D and 3D components of the flow, we focus on an important property of this decomposition, in any finite system: the decomposition implicitly depends on the vertical extent $\Delta h$ of the measurement volume. Increasing $\Delta h$ improves the resolution in $k_z$ and narrows the range of modes contributing to the $k_z \approx 0$ range. To quantify this effect, we compute the temporal spectra for different values of $\Delta h$, and define the crossover frequency $\omega^\star$ at which $E_{2D}(\omega^\star)=E_{3D}(\omega^\star)$.

\begin{figure}[t]
\centering
\includegraphics[width=0.9\textwidth]{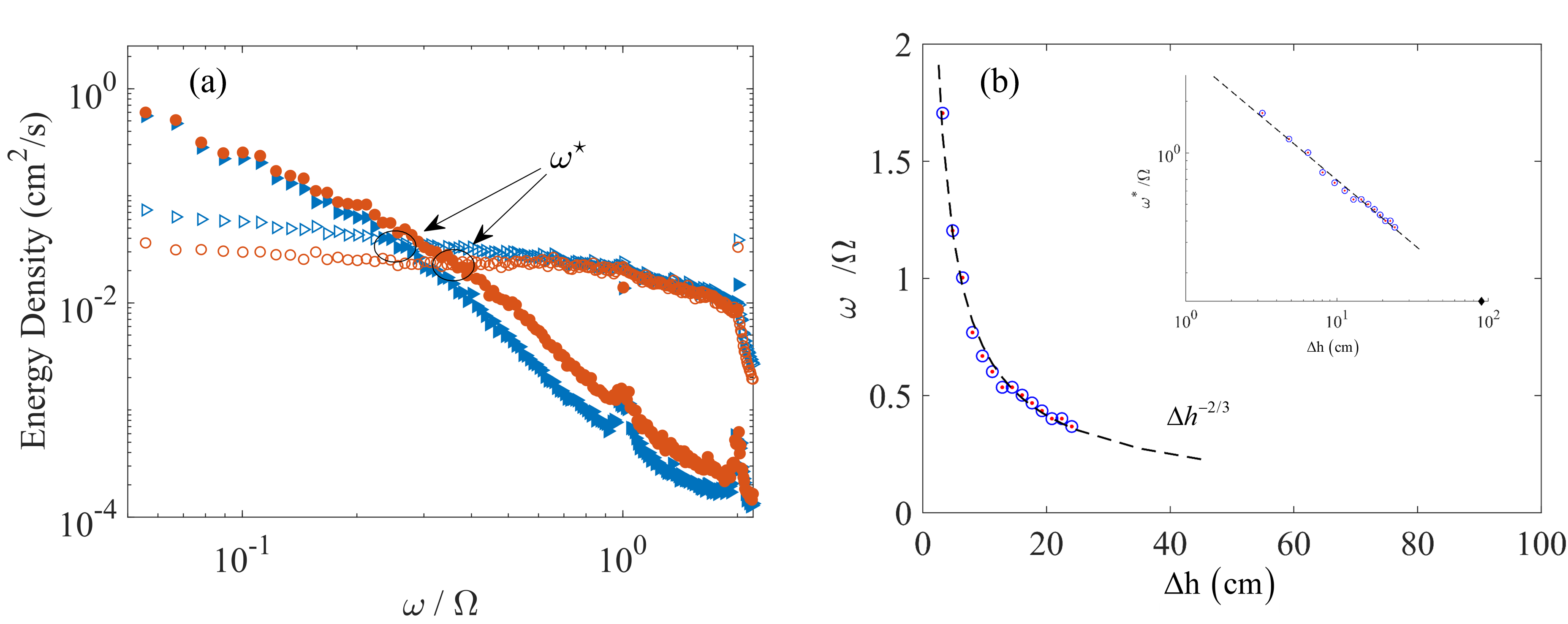}

\caption{
\textbf{(a)} Temporal energy spectra of the quasi-2D (solid symbols) and 3D (open symbols) velocity components for two values of the vertical scan size $\Delta h\approx 24 \, \mathrm{cm}$ and $\Delta h\approx 14 \, \mathrm{cm}$, illustrating the shift of the crossover frequency $\omega^\star$.
\textbf{(b)} Dependence of the crossover frequency $\omega^\star$ on $\Delta h$. The dashed line indicates a power-law fit over the accessible range of $\Delta h$.
}
\label{fig:omega_dh}
\end{figure}

Figure~\ref{fig:omega_dh} shows that the crossover frequency$\omega^\star$ decreases as $\Delta h$ increases. Over the accessible range of vertical scan sizes, the dependence of $\omega^\star$ on $\Delta h$ is approximately described by a power-law trend:
\begin{equation}
\omega^\star(\Delta h) \sim \Delta h^{-0.67}.
\end{equation}

In our experimental system the maximum value of $\Delta h$ is 24 cm. However, in order to estimate how influential the effect of vertical measurement range is, we extrapolate $\omega^\star(\Delta h)$ to the limit of $\Delta h \to L$, where $L=90cm$ is the system height, $\omega^\star$ shifted to substantially lower values, of order $0.2 \, \Omega$. This extrapolation is intended as an order-of-magnitude estimate illustrating the sensitivity of the decomposition to vertical resolution.

\subsection{Energy Partition Between Quasi-2D and 3D Components}

\begin{figure}
    \centering
    \includegraphics[width=8cm]{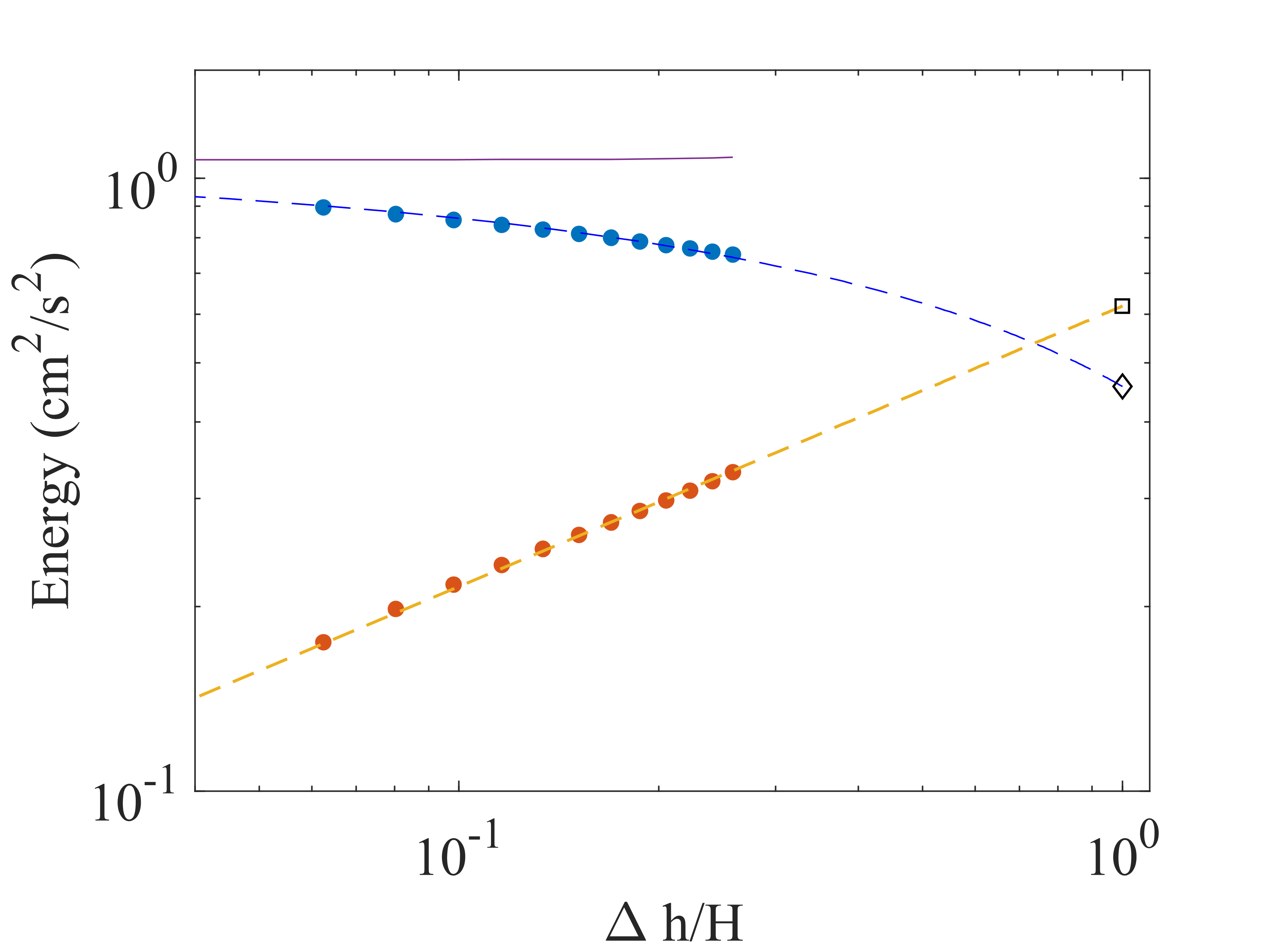}
    \caption{Total energy of the full velocity field (purple), quasi-2D flow (blue), and 3D field (orange), as a function of vertical scan height $\Delta h$. Dashed lines indicate extrapolations to the full system height.}
    \label{fig: total energy both parts with h}
\end{figure}

The dependence of the flow separation on $\Delta h$, affects the energy content in each component. As shown in Figure~\ref{fig: total energy both parts with h}, at small values of $\Delta h$, the quasi-2D component accounts for most of the measured energy. As $\Delta h$ increases,  the fraction of energy attributed to the quasi-2D component decreases while the energy associated with the 3D component increases. This trend reflects the improved resolution of vertical wavenumbers and the corresponding reclassification of modes with small but finite $k_z$. \OS{Although the classification changes with $\Delta h$, the avaraged large-scale component retained in the vertically averaged field still exhibits quasi-2D signatures, consistent with earlier rotating-turbulence measurements based on a single planar (2D) velocity fields \cite{yarom2013experimental}.} Extrapolating the measured trend to the full system height suggests that the 3D wave dominated field accounts for a significant fraction of the total energy. As in the previous subsection, this extrapolation is intended as an illustrative, order-of-magnitude estimate.

These measurements demonstrate that the apparent partition of energy between quasi-2D and three-dimensional motions depends explicitly on the vertical resolution of the measurement.


\section{Conclusions}

This study challenges the conventional view that rotating turbulence consists of independent 2D and inertial wave-driven three-dimensional flows. Through high-resolution volumetric measurements and careful spectral decomposition, we reveal that a separation between these flow components is strongly influenced by the finite vertical resolution of the experiments. 
The wave 3D field contains more and more energy at small frequencies and therefore, quasi-2D modes, modes with very small $k_z$, still obey the dispersion relation of inertial waves. With large enough vertical measurement range (high enough resolution in $k_z$), these modes are "classified" as 3D modes. Still, the flow they generate, presents the statistics which is characteristic of 2D turbulence. Naturally, as the vertical measurement range increases the ratio between the total energy in the 3D and the 2D components increases. Although not accessible in our present experimental setup, we cannot exclude the possibility that measurements spanning the full height of the tank would reveal a quasi-2D component that is not dominant, but instead comparable in energy to the three-dimensional wave component.

\OS{By resolving the temporal spectra of each component, we further show that the low-frequency, large-scale motions traditionally identified as quasi-2D cannot be cleanly separated from inertial-wave contributions in finite-window measurements.}
 The reported temporal scaling $E(\omega)\propto\omega^{-4/3}$ reported in \cite{yarom2014experimental} does not correspond to a distinct dynamical regime, but instead emerges from the superposition of slow quasi-2D motions and faster wave-dominated fluctuations. Its appearance and apparent exponent depend on how these contributions are mixed by the finite vertical measurement window.

These results call for a reassessment of theoretical descriptions that treat rotating turbulence as a superposition of decoupled two-dimensional and wave fields. Any realistic framework must account explicitly for the near-singular behavior as $k_z \to 0$ and for the resolution-dependent classification of modes. Such effects are unavoidable in numerical simulations and laboratory experiments, as well as in geophysical and astrophysical flows, where observational windows are necessarily finite.

\section*{Acknowledgments}
This research was supported by the Israel Science Foundation Grant $\#$ 2437/20.
\bibliographystyle{unsrt}

\bibliography{main_bib}

\end{document}